\documentclass[3p,times]{elsarticle}
\usepackage[T1]{fontenc}     
\usepackage[latin1]{inputenc}
\usepackage{lmodern}         
\usepackage{amsmath, amssymb, bbold, slashed, braket, siunitx}
\usepackage{float}
\usepackage{hyperref}
\usepackage{booktabs}
\usepackage{multirow}
\usepackage{amsthm}                       
\newtheorem{theorem}{Theorem}             
\newtheorem{corollary}{Corollary}[theorem]
\newtheorem{lemma}[theorem]{Lemma}        
\def\I{\mathbb{1}}                        
\newcommand{\pa}{\partial}                
\newcommand{\mna}{\left\langle{}m_{\nu_\alpha}\right\rangle}
\newcommand{\mne}{\left\langle{}m_{\nu_e}\right\rangle}     
\newcommand{\mnm}{\left\langle{}m_{\nu_\mu}\right\rangle}   
\newcommand{\mnt}{\left\langle{}m_{\nu_\tau}\right\rangle}  
\newcommand{\mnusup}{m_{\nu_{\mathrm{sup}}}}             
\newcommand{\anusup}{\alpha_{\nu_{\mathrm{sup}}}}        
\newcommand{\mnumax}{m_{\nu_{\mathrm{max}}}}             
\newcommand{\U}{{\scriptscriptstyle{\mathrm{U}}}}        
\newcommand{\D}{{\scriptscriptstyle{\mathrm{D}}}}        
\newcommand{\E}{{\scriptscriptstyle{\mathrm{E}}}}        
\newcommand{\Q}{{\scriptscriptstyle{\mathrm{Q}}}}        
\newcommand{\anu}{\alpha_\nu}                            
\newcommand{\dcp}{\delta_{\scriptscriptstyle\mathrm{CP}}}
\newcommand{\Rexp}{R_{\mathrm{exp}}}                     
\usepackage{makecell}
\setcellgapes{2pt}   

\begin{document}

\begin{frontmatter}

\title{\boldmath Massive Dirac neutrinos from an alternative extension of the Standard Model}

\author{M.A. De Andrade}
\ead{marco@fat.uerj.br}

\author{C. Neves}
\ead{clifford@fat.uerj.br}

\author{E.V. Corrêa Silva}
\ead{eduardo.vasquez@pq.cnpq.br}

\address{Departamento de Matemática, Física e Computação, \\
Faculdade de Tecnologia, \\ 
Universidade do Estado do Rio de Janeiro,\\
Rodovia Presidente Dutra, Km 298, Pólo Industrial,\\
CEP 27537-000, Resende-RJ, Brazil.}

\begin{abstract}
An alternative approach for deriving neutrino masses from the known masses of quarks, charged leptons, and the parameters of the Pontecorvo--Maki--Nakagawa--Sakata (PMNS) matrix is proposed. This framework is based on the simultaneous diagonalization of the kinetic and mass terms in an extension of the Standard Model (SM) with massive Dirac neutrinos, which have four degrees of freedom ($\nu_L$, $\nu_R$, and their antiparticles). Consistent with experimental data, numerical results for the neutrino masses are obtained through a calibration procedure that predicts the value of the CP phase $\dcp$, thereby reducing the number of independent flavor parameters in this SM extension. A scale parameter $\anu$ is introduced to adjust the mass splittings $\Delta m^2_{21}$ and $\Delta m^2_{31}$ to match their precise experimental values. The proposed framework connects neutrino mass generation to the known masses of quarks and charged leptons, offering a unified perspective on the matter sector of the SM.
\end{abstract}

\end{frontmatter}

\section{Introduction}
\label{sec:Intro}

At the turn of the 20th to the 21st century, a series of experiments confirmed the phenomenon of neutrino oscillations~\cite{pontecorvo1958,pontecorvo1968,
ohlsson2016,bilenky2016,kajita2016}. As a result, it is now widely accepted that the SM must be extended to include neutrino eigenstates, each with well-defined and distinct masses (mass eigenstates). This extension explains the existence of neutrino flavor states, which exhibit oscillatory behavior.

Neutrinos are unique particles that are observed only at the moment of their creation in a source and later, when they may transmute upon interaction with a
detector. They arise from a specific linear combination of three distinct mass eigenstates, all produced in the source with approximately the same linear
momentum. As they propagate freely, their wave-like nature causes them to travel at different speeds due to their mass differences. Since these masses are much smaller than that of any charged fermion, their speeds, though not identical, remain very close to the speed of light.

There are three unique modes of wave superposition that are sensitive to weak interactions---each corresponding to a different charged lepton. The wave
functions of these modes, mediated by charged weak interaction bosons, collapse at the source or detector, leading to the materialization of particles. These
particles are identified as electron neutrinos, muon neutrinos, and tau neutrinos, which materialize alongside their corresponding charged leptons.
This is a purely quantum phenomenon. Consequently, during wave propagation, it is physically meaningless to refer to electron, muon, or tau neutrinos. The neutrinos
associated with charged leptons do not propagate; they are only created in the source or materialized in the detector.

As a neutrino propagates, the evolution of its wave superposition over distance yields a non-zero probability of detecting a mode corresponding to a charged lepton different from the one initially produced at the source. This phenomenon underlies neutrino oscillations.

In the conventional extension of the SM with three right-handed neutrinos where total lepton number conservation is imposed~\cite{bilenky1980,bilenky1987} (which is the one with three massive Dirac neutrinos), seven additional physical parameters are introduced. These can be chosen as follows: one neutrino mass scale ($m_1$), the two mass splittings ($\Delta m^{2}_{21}$, $\Delta m^{2}_{31}$), the three mixing angles ($\theta_{12}$, $\theta_{13}$, $\theta_{23}$) and one CP phase ($\dcp$) from the PMNS matrix~\cite{pontecorvo1958,pontecorvo1968,maki1962,gribov1969}.

This is in addition to the nine fermion masses (three each for charged leptons, up quarks, and down quarks), and the four Cabibbo--Kobayashi--Maskawa (CKM) parameters (three mixing angles and one phase).

Therefore, the SM extended with three massive Dirac neutrinos contains $7 + 9 + 4 = 20$ flavor parameters.

The proposed extension introduces a new basis, called the twisted basis, in which the Lagrangian density features Dirac neutrino fields. In the kinetic term, the twisted neutrino states couple via a dimensionless, positive-definite Hermitian matrix $K$, constructed from the ratios of the quark Yukawa couplings. In the Yukawa interaction terms, they couple via a Hermitian mass matrix $M$, formed by the product of the charged-lepton Yukawa couplings and the Higgs boson field, transformed by the unitary PMNS matrix. The charged-fermion masses and the neutrino masses emerge from the standard Higgs interactions and subsequent Spontaneous Symmetry Breaking (SSB). Consequently, our extension operates entirely within the established SM. After returning to the mass basis, the complete spectrum becomes apparent.

In this framework, the independent flavor parameters in the neutrino sector are:
one dimensionless neutrino scale $\anu$ (which replaces the neutrino mass scale $m_1$), the two mass splittings ($\Delta m^{2}_{21}$, $\Delta m^{2}_{31}$), the three leptonic mixing angles ($\theta_{12}$, $\theta_{13}$, $\theta_{23}$), totaling six parameters---compared to seven in the standard extension. The CP phase $\dcp$ and the absolute neutrino mass scale $m_1$ are predictions of the model, and the inverted mass ordering (IO) is excluded because the condition $R(\dcp) = \Rexp$ has no solution for the IO CP phase within the full interval $[0^\circ, 360^\circ]$, as verified by a numerical scan using the NuFIT~6.1 best-fit values, which is the same numerical scan used to determine $\dcp$ with the parameters of normal mass ordering (NO).

In addition to the nine charged-fermion masses and the four CKM parameters, the proposed extension contains $6 + 9 + 4 = 19$ independent flavor parameters. This reduction from twenty to nineteen is achieved by removing the independent phase $\dcp$, which becomes a falsifiable prediction of the model. This will be demonstrated in Subsection \ref{sec:calibration}.

The article is organized as follows. 
In section~\ref{sec:ExtSpinorField} the matrix notation for the extended fermionic fields that will be utilized is introduced. 
In section~\ref{sec:SMmatter}, the matter sector of the SM of fundamental interactions, which consists of fermionic particles whose fields are organized
into extended spinors, is described. We also present the PMNS mixing matrix, which connects the basis of neutrino mass eigenstates to the basis of neutrino
flavor states that directly participate in weak interactions.
In section~\ref{sec:Diagonal}, an alternative extension of the SM is proposed and it allows to obtain Dirac neutrino masses from the simultaneous diagonalization of
matrices present in a twisted Lagrangian density of the neutrino matter sector. As far as we are aware, this procedure has not been previously reported in the literature.
In section~\ref{sec:Mass}, numerical results for neutrino masses are presented, such that the mass splittings $\Delta m^2_{21}$ and $\Delta m^2_{31}$ are adjusted in order to reach their precise experimental values. Our conclusions and outlook are given in section~\ref{sec:Conclusion}. 
\ref{sec:A.Lin} develops the necessary linear algebra tools for the simultaneous diagonalization of the kinetic and mass matrices of the twisted
Lagrangian density of the neutrino matter sector, aiming to reveal the mass content of the neutrino eigenstates.
\ref{sec:Cardano} outlines the analytical solution to the eigenvalue problem, which yields the neutrino masses.

\section{The extended spinor field}
\label{sec:ExtSpinorField}

Let the flavor index $a = 1, \dots, N$ denote the $N$ members of a family of fields. In the case of a $D$-dimensional or $(D+1)$-dimensional spacetime Dirac spinor, where $D$ is an even dimension, each member of the family is represented by a column vector with $n = 2^{D/2}$ components. For the four-dimensional spacetime spinor field, $n = 4$. Let $\I_n$ denote the $n \times n$ identity matrix. The Dirac Lagrangian density for a family of fields can be generalized as
\begin{equation}
\mathcal{L}=\overline\psi_a\left(K_{ab}\,i\slashed{\pa}-M_{ab}\,\I_n\right)\psi_b, \label{Lspinorcomp}
\end{equation}
where $\overline{\psi}_a = \psi_a^\dag{}\gamma^0$, $\slashed{\pa} = \gamma^\mu\pa_\mu$ and $\hbar = c = 1$. The parameters $K_{ab}$ are dimensionless, while $M_{ab}$ have units of mass. 

To construct the extended spinor field, we collect all family members $\psi_a$ into a single column vector $\psi$ with $Nn$ components. By utilizing the Kronecker product, the Lagrangian density can be compactly rewritten so that the kinetic and mass matrices, $K$ and $M$, act properly on the extended spinor space. Thus, the generalized Dirac Lagrangian density for the extended spinor field becomes
\begin{equation}
\mathcal{L} = \overline\psi\left(K\otimes{}i\slashed{\pa}-M\otimes{}\I_n\right)\psi, \label{Lspinor}
\end{equation} 
where
\begin{equation}
\overline\psi\equiv\psi^\dag\bigl(\I_N\otimes\gamma^0\bigr).  \label{barpsi}
\end{equation}
For the Lagrangian density to be real (in addition to being Lorentz invariant), the matrices $K$ and $M$ must be Hermitian. Consequently, these matrices together have at most $2N^2$ independent parameters. Our goal is to demonstrate that all these parameters can be combined into just $N$ mass parameters, corresponding to the masses of each family member.

Let us employ three equivalent bases to describe neutrinos:
\begin{itemize}
\item The \textit{flavor basis} $\nu_f$, which describes neutrinos as states participating in weak interactions alongside their corresponding charged leptons.
\item The \textit{mass basis} $\nu_m$, which describes neutrinos as eigenstates with well-defined masses that propagate freely.
\item The \textit{twisted basis} $\nu_t$, which describes neutrinos as states appearing in the twisted Lagrangian density of the neutrino sector, characterized by non-canonical kinetic and mass matrices.
\end{itemize}

\section{The Matter Sector of the Standard Model and the PMNS Matrix}
\label{sec:SMmatter}

The matter sector of the SM consists of kinetic and mass terms for four families of Dirac fermions, each with three flavor generations ($N = 3$) and four-component spinor fields ($n = 4$). Extended spinor fields with 12 components are constructed for each family, which can be separated into chiral components using the left and right chiral projection operators:
\begin{equation}
P_L=\frac12\left(\I_4 - \gamma_5\right), \quad P_R=\frac12\left(\I_4 + \gamma_5\right).
\label{eq:chiralproj}
\end{equation}
The projections for the charged lepton are given by
\begin{equation}
{\ell_f}_L = (\I_3 \otimes P_L) \, \ell_f, \quad {\ell_f}_R = (\I_3 \otimes P_R) \, \ell_f,
\end{equation}
and similarly for the other fields. For the charged-leptons and neutrino flavor states, we write
\begin{equation}
\ell_f = {\ell_f}_L + {\ell_f}_R = \begin{pmatrix}     e \\     \mu \\     \tau \end{pmatrix}, \quad
\nu_f  = {\nu_f}_L + {\nu_f}_R = \begin{pmatrix} \nu_e \\ \nu_\mu \\ \nu_\tau \end{pmatrix},
\label{psileptons}
\end{equation}
where $\nu_e$, $\nu_\mu$, and $\nu_\tau$ denote the electron, muon, and tau neutrinos, respectively. Similarly, the quark families are decomposed as
\begin{equation}
q^\U = q^\U_L + q^\U_R = \begin{pmatrix} u \\ c \\ t \end{pmatrix}, \quad
q^\D = q^\D_L + q^\D_R = \begin{pmatrix} d \\ s \\ b \end{pmatrix}.
\label{psiquarks}
\end{equation}
These extended spinors are usually associated with Dirac Lagrangian densities, which are derived from Eq.~\eqref{Lspinor} by setting $K = \I_3$ and choosing $M$ as a diagonal matrix whose elements correspond to the fermion masses in the family. However, within the SM, the neutrino flavor states remain massless.

\subsection{The PMNS Mixing Matrix}
\label{sec:PMNS}

Each left-handed neutrino flavor state in $\nu_f$ couples to its corresponding left-handed charged lepton in $\ell_f$ through the charged weak interaction bosons. Assuming that neutrinos also undergo Yukawa interactions, each left-handed neutrino mass eigenstate in $\nu_m$ couples to its corresponding right-handed counterpart through the Higgs field. Upon SSB, these eigenstates acquire their masses, propagate freely, and are assembled into components of the extended spinor
\begin{equation}
\nu_m = \begin{pmatrix} \nu_1 \\ \nu_2 \\ \nu_3 \end{pmatrix}.
\label{neutrinomasseigen}
\end{equation}
The PMNS mixing matrix $U$ relates the mass eigenstates to the flavor states,
\begin{equation}
\nu_f = \left(U \otimes \I_4\right) \nu_m,
\label{nuellnumu}
\end{equation}
so that each neutrino flavor state is a quantum superposition of all neutrino mass eigenstates.
The PMNS matrix is expressed as
\begin{equation}
U = \begin{pmatrix}
U_{e1}     &  U_{e2}     &   U_{e3}   \\
U_{\mu1}   &  U_{\mu2}   &   U_{\mu3} \\
U_{\tau1}  &  U_{\tau2}  &   U_{\tau3} 
\end{pmatrix}.
\label{PMNS}
\end{equation}
Consequently, the relation in Eq.~\eqref{nuellnumu} takes the standard component form:
\begin{equation}
\nu_\alpha = \sum_i U_{\alpha i} \, \nu_i,
\label{nuellnumuc}
\end{equation}
where $\alpha = e, \mu, \tau$ denotes the flavor index, and $i = 1, 2, 3$ labels the mass eigenstates.

A general $3 \times 3$ unitary matrix has nine degrees of freedom. However, in the case of the PMNS matrix $U$, five of these parameters can be absorbed as phases of lepton fields, leaving four independent real parameters. The mixing matrix is commonly parametrized as the product of three rotation matrices, using the Tait-Bryan parametrization \cite{chau1984} with mixing angles $\theta_{12}$, $\theta_{23}$, $\theta_{13}$, and one CP phase $\dcp$:
\begin{align}
&U=U_1\,U_2\,U_3,
\qquad\qquad\qquad
U_1=
\arraycolsep=5pt\def\arraystretch{1.3}
\begin{pmatrix}
1  &  0                  &  0                 \\
0  &  \cos(\theta_{23})  &  \sin(\theta_{23}) \\
0  &  -\sin(\theta_{23}) &  \cos(\theta_{23})
\end{pmatrix},
\quad \nonumber\\
&U_2=
\arraycolsep=5pt\def\arraystretch{1.3}
\begin{pmatrix}
\cos(\theta_{13})              &  0  &   \sin(\theta_{13})\,e^{-i\dcp} \\
0                              &  1  &              0                  \\
-\sin(\theta_{13})\,e^{i\dcp}  &  0  &   \cos(\theta_{13}) 
\end{pmatrix},  
\quad
U_3=
\arraycolsep=5pt\def\arraystretch{1.3}
\begin{pmatrix}
\cos(\theta_{12})   &  \sin(\theta_{12})  &   0 \\
-\sin(\theta_{12})  &  \cos(\theta_{12})  &   0 \\
0                   &  0                  &   1 
\end{pmatrix}.
\label{mixing_matrix}
\end{align}
For Majorana neutrinos, two additional phase angles $\eta_1$ and $\eta_2$ are introduced via the diagonal matrix $\mathcal{P} = \operatorname{diag}\left(e^{i\eta_1}, e^{i\eta_2}, 1\right)$, resulting in the mixing matrix $ U = U_1\,U_2\,U_3\,\mathcal{P}$ as described in \cite{gonzalez-garcia2008} and the standard Particle Data Group review \cite{PDG2024,PDG2026}. For Dirac neutrinos, $\mathcal{P}$ is the identity matrix, and the mixing matrix is the one given in Eq.~\eqref{mixing_matrix}.

\section{Physical requirements for the neutrino Lagrangian density}
\label{sec:Diagonal}

The physical requirements of a fundamental quantum field theory (QFT) are:
\begin{itemize}
	
\item A physical theory must agree with experiment.
	
\item A physical theory must be local such that its mathematical framework involves algebraic structures associated with spacetime regions, satisfying conditions such as microcausality and covariance. This ensures that physical influences propagate only within an object's immediate surroundings, excluding instantaneous action at a distance.
	
\item A physical theory must be unitary to preserve probability norms under time evolution. A key requirement for unitarity is that the Lagrangian density must be real.
	
\item The Lagrangian density of a fundamental physical theory must be invariant under Poincaré transformations, CPT-symmetric, and renormalizable.
	
\item Gauge invariance requires $\sum Q = \sum I_3 = \sum Y = 0$ for each term in the Lagrangian density, where $Q$, $I_3$, and $Y$ denote electric charge, weak isospin, and hypercharge, respectively.
	
\end{itemize}

While mathematical freedom in a theory should be fully explored, any application in theoretical physics must satisfy physical constraints that reduce the initial degrees of freedom offered by mathematics.

To generate Dirac masses for the Lagrangian density, we employ the standard $\mathrm{SU}(2)_L \times \mathrm{U}(1)_Y$ gauge-invariant Yukawa couplings. These couple the right-handed singlet fermions to the left-handed doublet via the Higgs doublet $\Phi$ for charged leptons and via its conjugate $\widetilde{\Phi} = i\tau_2\Phi^\ast$ for neutrinos. In this way, charged leptons and neutrinos acquire Dirac masses through their interaction with the Higgs field during SSB.

Using our notation, the Dirac Lagrangian density in the flavor basis for the matter sector of charged leptons that satisfy all the physical requirements is expressed  in the standard diagonal form as
\begin{equation}
\mathcal{L}_\ell=\overline\ell_f\left(\I_3\otimes{}i\slashed{\pa}-m_\E\otimes{}\I_4\right)\ell_f, \label{Laglepton}
\end{equation}
where $m_\E = \frac1{\sqrt{2}}\,v\,y_\E$ is the matrix whose elements are the masses of the electron, muon and tau,
\begin{equation}
m_\E= \begin{pmatrix}
m_e  &   0      &   0      \\
0    &   m_\mu  &   0      \\
0    &   0      &   m_\tau 
\end{pmatrix}. \label{leptonmass}
\end{equation}
Here $v \approx 246.22\,\text{GeV}$ denotes the Higgs vacuum expectation value, and $y_\E$ represents the diagonal Yukawa coupling matrix.

This extension of the SM differs from the conventional approach by introducing a new basis, the twisted basis, where the neutrino fields are coupled through parametrized kinetic and mass terms within the Lagrangian density. After a non-unitary equivalence transformation from the twisted basis to the mass basis, the Lagrangian density reduces to the conventional Dirac Lagrangian density with canonical kinetic and mass terms, causing certain parameters to become intertwined and lose their independence.

To maximize mathematical freedom in the twisted basis, the neutrino Lagrangian density includes a kinetic matrix and a Yukawa matrix. The latter couples the neutrino fields to the standard Higgs boson, so that the SSB of the gauge group $\mathrm{SU}(3)_\mathrm{C} \times \mathrm{SU}(2)_L \times \mathrm{U}(1)_Y$ induces neutrino interactions via the Higgs boson.

Constrained by the requirements of a fundamental QFT, the SSB generates twisted neutrino fields $\nu_t$ that couple to both a Hermitian positive-definite kinetic matrix $K$ (derived from a specific combination of quark Yukawa couplings) and a Hermitian positive-definite mass matrix $M$ (constructed from charged-lepton masses). These terms collectively generate neutrino masses after a non-unitary transformation from the twisted basis $\nu_t$ to the mass basis $\nu_m$, resulting in the standard diagonal form of a Dirac Lagrangian density, like the one in Eq.~\eqref{Laglepton}.

The weak interaction sector comprises the charged-lepton fields $\ell_{f}$ and their neutrino flavor counterparts $\nu_{f}$ (defined in Eq.~\eqref{psileptons}), whose dynamics are mediated by the charged gauge bosons $W^\pm_\mu$. Since this interaction acts exclusively upon the left-handed chirality sector of the Dirac spinor---leaving only left-handed Weyl spinors dynamically active---the left chiral projection operator $P_{L}$ from Eq.~\eqref{eq:chiralproj} must be used, yielding the neutrino Lagrangian density expressed as
\begin{equation}
\mathcal{L}_\nu=\overline\nu_t\left(K\otimes{}i\slashed{\pa}-M\otimes{}\I_4\right)\nu_t
-\frac{g}{\sqrt{2}}\,\overline\ell_f\left(\I_3\otimes\slashed{W}^{-}P_L\right)\nu_f+\text{h.c.}, 
\label{Lag}
\end{equation}
where $\slashed{W}^\pm = \gamma^\mu W^\pm_\mu$, $g$ is the $\mathrm{SU}(2)_L$ group's coupling constant, and ``h.c.'' denotes the Hermitian conjugate of the weak interaction term.

Our purpose is the simultaneous diagonalization of the kinetic matrix $K$ and mass matrix $M$, transforming $K$ to $\I_3$ and $M$ to a diagonal matrix $m_\nu$, whose elements correspond to the masses of the neutrino mass eigenstates:
\begin{equation}
m_\nu=\begin{pmatrix}
m_1  &   0    &   0      \\
0    &   m_2  &   0      \\
0    &   0    &   m_3 
\end{pmatrix}.   \label{massneutrinos}
\end{equation}
We analyze the reparametrization of the neutrino fields due to the simultaneous diagonalization of $K$ and $M$ and its impact on the neutrino sector of the Lagrangian density. Let $\Omega_\nu$ be the matrix responsible for the twisting of the neutrino matter sector, such that the mass eigenstates transform to twisted states as
\begin{equation}
\nu_t=\left(\Omega_\nu\otimes{}\I_4\right)\nu_m\quad\Rightarrow\quad\overline\nu_t = \overline\nu_m\left(\Omega_\nu^\dag\otimes{}\I_4\right). \label{spinorfield}
\end{equation}
After this transformation, the matter sector of the neutrino Lagrangian density given in Eq.~\eqref{Lag} becomes
\begin{equation}
\mathcal{L}_\mathrm{matter}=\overline\nu_m\left(\Omega_\nu^\dag{}K\Omega_\nu\otimes{}i\slashed{\pa}-\Omega_\nu^\dag{}M\,\Omega_\nu\otimes{}\I_4\right)\nu_m. \label{transfLnu}
\end{equation}
We aim to express the neutrino matter sector in the standard diagonal Dirac form:
\begin{equation}
\mathcal{L}_\mathrm{matter} = \overline\nu_m\left(\I_3\otimes{}i\slashed{\pa}-m_\nu\otimes{}\I_4\right)\nu_m, \label{Lnumatter}
\end{equation}
where $m_\nu$ is given in Eq.~\eqref{massneutrinos}.
This requires solving the system of matrix equations,
\begin{equation}
\def\arraystretch{1.5}
\begin{cases}
\Omega_\nu^\dag{}K\Omega_\nu = \I_3, \\
\Omega_\nu^\dag{}M\,\Omega_\nu = m_\nu.
\end{cases}                                  
\label{originalsystem}
\end{equation}
Since $K$ is not the identity matrix, it follows from this system that $\Omega_\nu$ is non-unitary, as can be verified in detail in \ref{sec:A.Lin}. For the system~\eqref{originalsystem} to be recast in a solvable form, $K$ must necessarily be invertible. Henceforth, we assume this condition holds, which allows us to express the first equation of this system as
\begin{equation}
\Omega_\nu^{-1}K^{-1}{\Omega_\nu^\dag}^{-1}=\I_3.     \label{invfirstsystemeq}
\end{equation}
From this, we obtain
\begin{equation}
\def\arraystretch{1.5}
\begin{cases}
\Omega_\nu\Omega_\nu^\dag = K^{-1}, \\
\Omega_\nu^{-1}K^{-1}M\,\Omega_\nu = m_\nu.
\end{cases}
\label{modifiedsystem}
\end{equation}
The second equation of the system~\eqref{modifiedsystem} was derived by multiplying matrix expressions, where the first factor originates from Eq.~\eqref{invfirstsystemeq} and the second factor comes from the second equation of the system~\eqref{originalsystem}.

We solve the eigenvalue problem for system~\eqref{modifiedsystem} using the linear algebra framework from \ref{sec:A.Lin} (corollary~\ref{eigenvalueproblem}). Under the assumption that both $K^{-1}$ and $M$ are Hermitian positive-definite matrices, their product $K^{-1}M$, though non-Hermitian, is guaranteed to have strictly positive eigenvalues.  
Since the second equation from the system~\eqref{modifiedsystem} is a matrix similarity relation, these positive eigenvalues correspond to the masses $m_1$, $m_2$, and $m_3$ of the neutrino eigenstates, forming the diagonal matrix $m_\nu$ in Eq.~\eqref{massneutrinos}. While $K^{-1}M$ arises in linear algebraic analysis, it is essential to emphasize that this non-Hermitian product appears in \emph{no} Lagrangian density; all such densities exclusively contain Hermitian matrices.  
Finally, using Eqs.~\eqref{nuellnumu}, \eqref{spinorfield} and \eqref{originalsystem}, the Lagrangian density in Eq.~\eqref{Lag} is fully expressed in terms of the neutrino mass eigenstates, taking the standard form.
\begin{equation}
\mathcal{L}_\nu = \overline{\nu}_m\left(\I_3\otimes{}i\slashed{\pa}-m_\nu\otimes{}\I_4\right)\nu_m
-\frac{g}{\sqrt{2}}\,\overline\ell_f\left(U\otimes\slashed{W}^{-}P_L\right)\nu_m+\text{h.c.}    \label{Lagmassstate}
\end{equation}
The eigenstates $\nu_m$ propagate as plane waves with well-defined masses, but they are never observed individually due to the non-diagonal nature of the weak interaction term. Instead, what detectors measure are the flavor states---three distinct linear combinations of these mass eigenstates mixed by the PMNS matrix. These flavor states emerge after diagonalization of the weak interaction term and are identified as the electron ($\nu_e$), muon ($\nu_\mu$), and tau ($\nu_\tau$) flavor neutrinos, each associated with their corresponding charged lepton.

The neutrino Lagrangian density in Eq.~\eqref{Lagmassstate} can be expressed entirely in terms of flavor states through the inverse transformation of Eq.~\eqref{nuellnumu},
\begin{equation}
\nu_m=\left(U^\dag\otimes{}\I_4\right)\nu_f.
\end{equation}
This transformation leads to the definition of the Hermitian neutrino mass matrix in the flavor basis,
\begin{equation}
M_\nu=U\,m_\nu\,U^\dagger,
\end{equation}
which allows us to rewrite the Lagrangian density as
\begin{equation}
\mathcal{L}_\nu = \overline{\nu}_f\left(\I_3\otimes{}i\slashed{\pa}-M_\nu\otimes{}\I_4\right)\nu_f
-\frac{g}{\sqrt{2}}\,\overline\ell_f\left(\I_3\otimes\slashed{W}^{-}P_L\right)\nu_f+\text{h.c.} \label{Lagweakstate}
\end{equation}
When the Lagrangian density is expressed in terms of flavor states---the states that directly participate in weak interactions---the resulting neutrino mass matrix $M_\nu$ is non-diagonal. This non-diagonal structure implies that these flavor states do not propagate although they are the only ones that can be observed during weak interaction processes. The diagonalization of $M_\nu$ transforms the Lagrangian density back to the form given in Eq.~\eqref{Lagmassstate}, revealing the true masses of the neutrino eigenstates, although they cannot be individually observed.

\subsection{Phenomenological choice of the neutrino kinetic and mass matrices}
\label{sec:phenomelogy}

To relate neutrino masses to the known quark and charged-lepton masses, we construct the positive-definite Hermitian kinetic matrix using the known Yukawa couplings of the quark fields $q^\U$ and $q^\D$, while the Hermitian mass matrix is built from the known masses of the charged-lepton fields $\ell_f$, thereby interconnecting all particles within the matter sector of the SM.

The quark Yukawa couplings are combined into a dimensionless kinetic matrix, encoding the CKM structure that emerges from the biunitary diagonalization of the quark Yukawa coupling matrices. This kinetic matrix is dynamically linked to the mass matrix, which is constructed from the charged-lepton masses and transformed via the PMNS mixing matrix.

Introducing the matrices $m_\U=\frac1{\sqrt{2}}\,v\,y_\U$, whose elements are the masses of the up, charm, and top quarks, and $m_\D=\frac1{\sqrt{2}}\,v\,y_\D$, whose elements are the masses of the down, strange, and bottom quarks,
\begin{equation}
m_\U=\begin{pmatrix}
m_u  &   0    &   0      \\
0    &   m_c  &   0      \\
0    &   0    &   m_t 
\end{pmatrix}, \quad
m_\D=\begin{pmatrix}
m_d  &   0    &   0      \\
0    &   m_s  &   0      \\
0    &   0    &   m_b 
\end{pmatrix}.  \label{quarks}
\end{equation}
Here, $y_\U$ and $y_\D$ represent the diagonal Yukawa coupling matrices.
We define the Yukawa matrices $Y_\U$ and $Y_\D$ through the biunitary transformations
\begin{equation}
Y_\U = U_u\,y_\U\,V_u^\dag, \quad
Y_\D = U_d\,y_\D\,V_d^\dag,    
\label{biunitary}
\end{equation}
such that the matrices $U_u$ and $U_d$ are related to the CKM mixing matrix $V$ through the physical condition
\begin{equation}
V = U_u^\dag \, U_d.
\label{CKM}
\end{equation}
Taking into account these standard biunitary transformations, and assuming Hermiticity under specific constraints in our procedure, we propose a definition for $K$ and $M$ to simultaneously account for the appropriate CP phase and the experimental mass splittings. Consequently, the unique dimensionless kinetic matrix $K$ that survives---without introducing any additional parameters beyond those of the Standard Model---is given by
\begin{equation}
K = Y_\U Y_\D^{-1} V.
\label{_kinetic}
\end{equation}
Numerical analysis indicates that alternative choices for $K$ fail to simultaneously sustain both requirements. The similarity-transformed matrix $(V K V^\dagger = V Y_\U Y_\D^{-1})$ can be excluded from the discussion as it yields identical results. The Hermiticity requirement for $K$ constrains the unphysical matrices $V_u$ and $V_d$ in the biunitary transformations, uniquely determining the matrix $U_u$ as
\begin{equation}
V_d = V_u, \quad U_u = \I_3.   
\label{specbiuni}
\end{equation}
The requirement $U_u = \I_3$ leads to $U_d = V$ without imposing any restriction on the CKM matrix. The condition $V_d = V_u$ poses no problem since, in the SM, the right-handed mixing matrices $V_u$ and $V_d$ from the biunitary transformations (Eq.~\eqref{biunitary}) are \textit{unphysical}. This holds because:
\begin{itemize}
\item Right-handed quarks are singlets under $\mathrm{SU(2)}_L$ and do not participate in weak interactions.
\item The $V_u$ and $V_d$ rotations can be absorbed through unobservable redefinitions of the right-handed quark fields.
\item Only the combination $U_u^\dag U_d$ (the CKM matrix) is physically observable.
\end{itemize}
Thus, substituting Eqs.~\eqref{biunitary} and \eqref{CKM} into Eq.~\eqref{_kinetic} and applying the conditions given in Eq.~\eqref{specbiuni}, we obtain the positive-definite Hermitian kinetic matrix $K$ as
\begin{equation}
K=y_\U\,y_\D^{-1}=m_\U\,m_\D^{-1}.
\label{kinetic}
\end{equation}

Our objective is to intertwine the Yukawa couplings of all charged leptons and quarks to generate light neutrino Yukawa couplings consistent with experimental data. Another reason for this structure is that the mass matrix $M$, which contains the PMNS matrix $U$ in Eq.~\eqref{Mell}, yields---together with the inverse matrix $K^{-1}$---an expression that may be transformed via arbitrary unitary rotations of $K$ and $M$ in such a way that the transformed product ${K'}^{-1}M'$ will always produce the same mass eigenvalues. Thus, the light neutrino masses will be invariant under unitary rotations of the matrices $K$ and $M$.
Thus, the positive-definite Hermitian mass matrix $M$ is defined via the similarity transformation
\begin{equation}
M = \anu\,U m_\E U^\dagger, 
\label{Mell}
\end{equation}
where $\anu$ is the neutrino scaling parameter, $U$ (Eq.~\eqref{mixing_matrix}) denotes the PMNS mixing matrix, and $m_\E$ (Eq.~\eqref{leptonmass}) is the diagonal charged-lepton mass matrix of the flavor sector. This definition introduces five quantities beyond the SM: the scale parameter $\anu$, the three PMNS mixing angles, and the CP phase $\dcp$.

Using the $K$ and $M$ matrices as given in Eqs.~\eqref{kinetic} and \eqref{Mell}, the matrix $K^{-1}M$ is expressed in terms of known physical quantities as
\begin{equation}
K^{-1}M=\anu\,m_\D\,m_\U^{-1}\,U\,m_\E\,U^\dag,
\label{invKMell}
\end{equation}
where, as can be seen from the similarity relation in the second equation of the system~\eqref{modifiedsystem}, the eigenvalues of $K^{-1}M$ yield the neutrino masses $m_i$ ($i = 1, 2, 3$).

Due to the novel role of the PMNS matrix in contributing directly to neutrino masses (as shown in Eq.~\eqref{invKMell}), we introduce the auxiliary unitary matrices $U_\Q$ and $U_\E$, which represent arbitrary rotations of the quark and lepton \textit{mass matrices}. These matrices are related to the PMNS matrix as
\begin{equation}
U = U_\Q^\dagger U_\E.  
\label{pmns_decomp}
\end{equation}

Although our notation resembles the matrices in the standard biunitary transformation $U = U_\ell^\dagger U_\nu$, our matrices $U_\Q$ and $U_\E$ play a fundamentally different role: they do not diagonalize the neutrino or charged-lepton mass terms, nor do they correspond to transformations of the quark and lepton fields.
Instead, they are matrices that merely parametrize the freedom to rotate the mass matrices $m_\U$, $m_\D$, and $m_\E$ used to construct $K$ and $M$, without affecting the physical neutrino masses. It is important to emphasize that the standard biunitary matrices ($U_\ell, V_\ell$ and $U_\nu, V_\nu$) that diagonalize the charged-lepton and neutrino mass matrices, respectively, are not replaced by $U_\Q$ and $U_\E$; they continue to exist as the physical mixing matrices.

Substituting Eq.~\eqref{pmns_decomp} into Eq.~\eqref{invKMell}, one finds
\begin{equation}
U_\Q K^{-1} M U_\Q^\dag = M_\D' (M_\U')^{-1} M' = K'^{-1} M',  
\label{transformed}
\end{equation}
with the transformed Hermitian matrices defined as
\begin{align}
M_\D' &= U_\Q m_\D U_\Q^\dag, \\ 
M_\U' &= U_\Q m_\U U_\Q^\dag, \\ 
M'    &= \anu\,U_\E m_\E U_\E^\dag.  
\label{primed_masses}
\end{align}
The matrix similarity relation between ${K'}^{-1}M'$ and $K^{-1}M$ in Eq.~\eqref{transformed} demonstrates that the coupling of neutrino twisted states in the Lagrangian density in Eq.~\eqref{Lag} is possible for any pair $(K', M')$ without affecting neutrino masses. This invariance property implies that our specific choice $K = m_\U m_\D^{-1}$ (i.e., $U_\Q = \I_3$) represents merely a convenient choice of unitary rotation. Any other choice related by a unitary rotation of the quark fields yields the same physics, provided that the corresponding $M$ is adjusted accordingly.  As a practical example, by exploiting the explicit form of $U$ given in Eq.~\eqref{mixing_matrix}, one can, devoid of any physical pretension, choose $U_\Q = U_1^\dagger$ and $U_\E = U_2\,U_3$.

\section{Results from this alternative approach}
\label{sec:Mass}

For our analysis, we present results obtained using the numerical data of physical quantities provided in \cite{PDG2026}. We adopt the most recent neutrino oscillation data from the \textit{IceCube/DeepCore 2024} (IC24) experiment \cite{abbasi2025} and the NuFIT~6.1 global fit \cite{NuFIT6.1}, yielding the values summarized in Table~\ref{PMNS_table}.
\begin{table}[H]
\centering
\makegapedcells
\caption{IC24 with SK atmospheric data---NuFIT~6.1 (2025).}
\label{PMNS_table}
\begin{tabular}{|l|c c|}
\hline
\multicolumn{1}{|c|}{~} & \multicolumn{2}{c|}{Normal Ordering (best fit)} \\
\cline{2-3}
\multicolumn{1}{|c|}{~} & \multicolumn{1}{c}{bfp $\pm 1\sigma$} & \multicolumn{1}{c|}{$3\sigma$ range}  \\
\hline
$\theta_{12}/^\circ$ & $33.76^{+0.42}_{-0.41}$  & $32.54 \rightarrow 35.03$ \\
$\theta_{23}/^\circ$ & $43.29^{+0.96}_{-0.79}$  & $41.27 \rightarrow 49.86$ \\
$\theta_{13}/^\circ$ & $8.62^{+0.11}_{-0.11}$   & $8.26 \rightarrow 8.95$ \\
$\dcp/^\circ$        & $212^{+26}_{-36}$        & $125 \rightarrow 365$ \\
$\Delta m_{21}^2 / (10^{-5}~\text{eV}^2)$ & $7.537^{+0.094}_{-0.10}$    & $7.236 \rightarrow 7.823$ \\
$\Delta m_{31}^2 / (10^{-3}~\text{eV}^2)$ & $+2.511^{+0.021}_{-0.020}$  & $+2.450 \rightarrow +2.576$ \\
\hline
\end{tabular}
\end{table}
\noindent The leptonic CP phase $\dcp$ accounts for CP symmetry violation in weak interactions.
The quantities that are available from neutrino oscillations experiments are the mass splittings, which are mapped analytically via:
\begin{align}
\Delta m_{21}^2 &= m_2^2 - m_1^2, \\
\Delta m_{31}^2 &= m_3^2 - m_1^2.
\end{align}
Thus, the physical masses of the individual neutrino mass eigenstates can be expressed as
\begin{equation}
m_1, 
\quad
m_2=\sqrt{m_1^2+\Delta m_{21}^2},
\quad
m_3=\sqrt{m_1^2+\Delta m_{31}^2}.
\label{masssplit}
\end{equation}

The extremal bounds on the neutrino mass scale $m_\nu$ are obtained by appropriately combining the charged-lepton mass eigenvalues. Consequently, the optimal quark-lepton combinations that yield these minimal and maximal bounds lead to the following physical inequality:
\begin{equation} 
\anu \left( \frac{m_b}{m_t} m_e \right) < m_\nu < \anu \left( \frac{m_d}{m_u} m_\tau \right).
\label{inequality}
\end{equation}
Isolating $\anu$ from the upper bound limit yields the structural relation:
\begin{equation} 
\anu \approx \frac{m_u}{m_d \, m_\tau} \mnumax.
\label{alpha_v_def}
\end{equation}
Adopting $\mnusup \equiv \SI{1}{eV}$ as the standard benchmark upper-limit neutrino mass scale, we define the corresponding reference scale parameter as
\begin{equation} 
\anusup = \frac{m_u}{m_d \, m_\tau} \mnusup.
\label{alpha_vsup}
\end{equation}
The dynamical neutrino scaling parameter is thus adjusted according to:
\begin{equation}
\anu = f \, \anusup,
\label{alpha_v_adjusted}
\end{equation}
where $f$ denotes an auxiliary dimensionless scaling parameter to be calculated later.

\subsection{Calibration procedure}
\label{sec:calibration}

The physical neutrino masses are obtained by adjusting two parameters in the matrix $K^{-1}M$, ensuring that its eigenvalues reproduce the experimental squared mass differences $\Delta m^2_{31}$ and $\Delta m^2_{21}$ listed in Table~\ref{PMNS_table}. The first parameter, the CP phase $\dcp$, is selected because the dimensionless ratio $R(\dcp) \equiv \Delta m^2_{31}/\Delta m^2_{21}$ depends exclusively on $\dcp$ (through the PMNS matrix $U(\dcp)$) and is independent of the overall mass scale, as derived in \ref{sec:Cardano}. A second, practical reason is that the large experimental uncertainty on $\dcp$ allows flexible adjustments within the currently allowed range.
The second parameter, the neutrino scaling factor $\anu$, is then adjusted to set the absolute mass scale without affecting $\dcp$.

The ratio $R(\dcp)$ is insensitive to variations in $\anu$, which motivates a two-step optimization. First, $\dcp$ is varied until the theoretical ratio matches the experimental value $R(\dcp) = \Rexp = 33.316$ (obtained from the data in Table~\ref{PMNS_table}), thereby predicting $\dcp$.

Then, the auxiliary parameter $f$ (or equivalently $\anu$) is adjusted to scale both $\Delta m^2_{31}$ and $\Delta m^2_{21}$ while keeping their ratio $\Rexp$ fixed, until both mass splittings agree with their best-fit experimental measured values. This step fixes the absolute mass scale $m_1$; the other masses $m_2$ and $m_3$ follow from the corresponding mass splittings, using Eq.~\eqref{masssplit}.

This framework excludes the inverted neutrino mass ordering (IO) because the condition $R(\dcp) = \Rexp$ admits no solution for the IO CP phase $\dcp$ within the full interval $[0^\circ, 360^\circ]$, as verified by the same numerical scan used to determine $\dcp$ with the parameters of the normal ordering (NO). Consequently, we adopt the PMNS parameters for the NO, defined by $m_1 < m_2 < m_3$. This exclusion of the IO provides an additional testable signature.

Since $\dcp$ and $m_1$ are predicted by the model, the new neutrino-independent parameters are $\anu$ (which replaces $m_1$), the two mass splittings, and the three mixing angles of the PMNS matrix, making a total of 6 independent parameters.

The numerical values for the three PMNS mixing angles are taken from NuFIT~6.1 global analysis showed in Table~\ref{PMNS_table}. Additionally, the six quark Yukawa couplings and the three charged-lepton masses are obtained from the 2026 Particle Data Group (PDG) review \cite{PDG2026}.

Propagation of uncertainties for all quantities uses the best-fit point ($\mathrm{bfp}$) as the central value, with uncertainty bounds defined by the $1\sigma$ endpoints of $\Delta m^2_{21}$ and $\Delta m^2_{31}$, i.e., $\mathrm{bfp} - \sigma_{\mathrm{lower}}$ and $\mathrm{bfp} + \sigma_{\mathrm{upper}}$. The optimization yields
\begin{equation}
\dcp = 230.6^\circ \pm 8.4^\circ, \quad f = 0.936 \pm 0.017,
\end{equation}
which corresponds to the neutrino scaling parameter
\begin{equation}
\anu = (2.421 \pm 0.064) \times 10^{-10}.
\end{equation}
Consequently, the physical extremal bounds in Eq.~\eqref{inequality} become
\begin{equation} 
3.00\times10^{-6}\,\,{eV} < m_\nu < {0.936}\,\,{eV}.
\label{inequality_with_numbers}
\end{equation}

Our central value $\dcp = \ang{230.6}$ is in good agreement with the allowed range quoted in Table~\ref{PMNS_table}. Using the PMNS parameters for normal ordering, the eigenvalues of $K^{-1}M$ give the following individual neutrino mass spectrum:
\begin{align}
m_1 &= (1.206 \pm 0.060) \times 10^{-5}\SI{}{eV},  \label{em1}
\\
m_2 &= (8.682 \pm 0.054) \times 10^{-3}\SI{}{eV},  \label{em2}
\\
m_3 &=  (5.011 \pm 0.021) \times 10^{-2}\SI{}{eV}.  \label{em3}
\end{align}
For completeness, the predicted neutrino Yukawa couplings ($y_\nu = \sqrt{2}\,m_\nu / v$), evaluated at the electroweak scale, are
\begin{align}
y_1 &\approx 6.9 \times 10^{-17},\nonumber\\
y_2 &\approx 5.0 \times 10^{-14},\\
y_3 &\approx 2.9 \times 10^{-13}.\nonumber
\end{align}
These extremely small values are a direct consequence of the model, not an \textit{ad hoc} assumption. The extreme smallness of the predicted Yukawa couplings is
technically natural in the 't Hooft sense \cite{thooft}. In the limit $y_\nu \rightarrow 0$, the SM extension Lagrangian density changes to the SM one and, as a consequence, it acquires an exact global $\mathrm{U(1)}_L$ lepton number symmetry. Consequently, all radiative corrections to the Yukawa couplings are proportional to the couplings themselves, yielding only a mild logarithmic running $\delta y_\nu \propto y_\nu \ln(\Lambda/M_W)$. For an ultraviolet cutoff as high as the Planck scale $\Lambda \sim M_{\rm Pl}$, we get $\ln(\Lambda/M_W) \approx 40$. Then the correction remains of order $y_\nu \times 40$, which does not destabilize the small values we obtain. This symmetry protection, combined with the renormalizability of the theory, guarantees that the tiny neutrino Yukawa couplings are not driven to $\mathcal{O}(1)$ by quantum effects, reinforcing the consistency of the model. Even if each coupling runs by a factor of 40 related to the Planck and electroweak  scales, the values remain in the range $10 ^{-15}$---$10^{-11}$, still negligibly small compared to any other SM Yukawa coupling. This demonstrates the viability of the predictions. Therefore, the small Dirac neutrino mass is perfectly stable against radiative corrections and the neutrinos play zero role in the hierarchy problem of the Higgs mass. As the tiny SM Yukawa couplings are technically natural and stable under quantum corrections, no fine-tuning from $\mathcal{O}(1)$ additive corrections or new physics is required to keep them small.

At the decoupling temperature $T \sim 1\,\text{MeV}$ \cite{hannestad1,hannestad2}, all three masses, given in 
Eqs.~\eqref{em1}--\eqref{em3}, satisfy $m_i \ll T$ for $i=1,2,3$. Therefore, at the temperature 
of neutrino decoupling, all three neutrino mass eigenstates are 
highly relativistic. Consequently, they contribute to $N_{\text{eff}}$ exactly 
as the three standard left-handed neutrinos. 

The right-handed components $\nu_R$ are completely sterile with respect to the SM gauge interactions and can only be populated through their Yukawa couplings 
$y_\nu$, which are of order $10^{-16}$ to $10^{-13}$. The resulting interaction rate is far too small to bring the right-handed states into thermal equilibrium at any temperature below the Planck scale. Hence, the $\nu_R$ fields do not contribute to $N_{\text{eff}}$, and the model is fully consistent with the cosmological bound $N_{\text{eff}} = 2.99±0.17$ \cite{Planck2018}.

Nothing prevents the neutrino from briefly acquiring a materialized mass at the moment of its creation or later at the instant of its detection. During nuclear beta decay, the effective flavor masses $\mna$ ($\alpha=e, \mu, \tau$) which are briefly observed at the source and detector, can be estimated using the absolute squares of the elements of the PMNS matrix, as well as the squares of the well-defined masses $m_i$ ($i=1, 2, 3$) of the neutrino mass eigenstates. By applying a weighted average formula \cite{PDG2024,weinheimer1999,vissani2001,weinheimer2005}, we determine $\mna$ via:
\begin{equation}
\mna \equiv \sqrt{\sum_i \left|U_{\alpha{}i}\right|^2 m_i^2}.
\end{equation}
By taking the absolute square of each element of the matrix given in Eq.~\eqref{PMNS} and the squares of the masses provided in Eqs.~\eqref{em1}, \eqref{em2} and \eqref{em3}, we obtain the following neutrino flavor masses,
\begin{align}
\mne &= (8.8972 \pm 0.0085) \times 10^{-3}\SI{}{eV},
\\
\mnm &= (3.443 \pm 0.012) \times 10^{-2}\SI{}{eV},
\\
\mnt &= (3.636 \pm 0.015) \times 10^{-2}\SI{}{eV}.
\end{align}
In the context of beta decay from tritium ($^3\mathrm{H}$) disintegration, experimental measurements set an upper limit for the electron neutrino mass at $\mne < \SI{0.45}{eV}$ \cite{PDG2026}. Similarly, the mass limits for the other flavors are $\mnm < \SI{190}{keV}$ and $\mnt < \SI{18.2}{MeV}$ \cite{PDG2026}. 

Quantitatively (as in \cite{NuFIT6.1} page~21), the global analysis of neutrino oscillation data together with the bound from the KATRIN experiment \cite{KATRIN2024} implies that at \SI{95}{\%} Confidence Level:
\begin{align}
\SI{0.00085}{eV} &\leq \mne \leq \SI{0.4}{eV} \quad \text{for normal ordering (NO)}, \\
\SI{0.058}{eV} &\leq \sum_\alpha\mna \leq \SI{1.2}{eV} \quad \text{for NO}.
\end{align}
Our results are consistent with these experimental bounds.

\subsection{Jarlskog invariant}

The strength of leptonic CP violation can be quantified by the Jarlskog invariant \cite{jarlskog1985}, 
defined using the PMNS parameters in Eq.~\eqref{mixing_matrix} as
\begin{equation}
J = \cos\theta_{12}\,\cos^2\theta_{13}\,\cos\theta_{23}\,\sin\theta_{12}\,\sin\theta_{13}\,\sin\theta_{23}\,\sin\dcp.
\label{eq:jarlskog_pmns}
\end{equation}
Using the current best-fit mixing angles from the recent PDG 2026 review \cite{PDG2026} and our predicted phase $\dcp = \ang{230.6}$, we obtain:
\begin{equation}
J = (-2.61 \pm 0.32) \times 10^{-2}.
\end{equation}
Setting the CP violating phase to its maximal value ($\dcp = \ang{270}$), the upper bound allowed exclusively by the experimental PMNS mixing angles is $|J_{\max}| \approx 0.0331$. Thus, our predicted invariant captures approximately {79}{\%} of the maximum physically available leptonic CP violation, demonstrating a large CP-violating effect, consistent with the global fit constraints \cite{NuFIT6.1}.

\subsection{Non-Unitary Twisting Matrix}

The twisting non-unitary matrix $\Omega_\nu$ evaluated at the best-fit point ($\mathrm{bfp}$) is explicitly calculated as
\begin{equation}
\arraycolsep=7pt\def\arraystretch{1.3}
\Omega_\nu = \begin{pmatrix}
0.2546                          &  0.7462                          &  1.2467 
\\
- 0.1100 + 0.02027 \,\mathrm{i} &  0.1418 + 0.1620 \,\mathrm{i}    & - 0.06238 - 0.1011 \,\mathrm{i} 
\\
0.1374 + 0.02193 \,\mathrm{i}   &  0.006234 + 0.05105 \,\mathrm{i} & - 0.03180 - 0.03503 \,\mathrm{i} 
\end{pmatrix}.
\end{equation}
Let ${\Omega_\nu}_{ij}$ denote each individual element of the matrix $\Omega_\nu$, and let \(\nu'_i\) represent the components of the extended twisted spinor \(\nu_t\). The algebraic expression relating the twisted basis to the mass basis (Eq.~\eqref{spinorfield}) can then be written as
\begin{equation}
\nu'_i = \sum_j {\Omega_\nu}_{ij} \, \nu_j.
\end{equation}

\section{Conclusion}
\label{sec:Conclusion}

In this work, we have proposed an extension of the Standard Model with Dirac neutrino masses, based on a new basis, called the twisted basis, which introduces non-canonical kinetic and mass matrices in the initial Lagrangian.  Through a non-unitary transformation $\Omega_\nu$, these matrices are simultaneously diagonalized, yielding the physical neutrino mass eigenstates in a basis where the kinetic term becomes canonical and the mass matrix diagonal.  All Lagrangian densities involved are physically equivalent, and in the mass basis the fundamental properties of unitarity, renormalisability, and gauge/CPT invariance are manifest.

One feature of this framework is the reduction of the independent flavor parameters. In the standard extension of the SM with three massive Dirac neutrinos, the neutrino sector requires seven independent parameters: the three PMNS angles, the CP phase $\dcp$, the absolute mass scale $m_1$, and the two mass squared differences. Adding the nine charged-fermion masses and the four CKM parameters results in a total of twenty independent parameters.
The calibration procedure then replaces $\dcp$ with the experimental mass splitting ratio $\Rexp = \Delta m^2_{31}/\Delta m^2_{21}$, which fixes $\dcp$ uniquely. The scale $\anu$ is subsequently adjusted to match one of the splittings, setting the absolute mass scale $m_1$, while $m_2$ and $m_3$ follow automatically from the corresponding mass splittings. 
As a result, the independent parameters in the neutrino sector become $\anu$, the three PMNS mixing angles, and the two mass splittings---six in number, but with $\dcp$ now a derived prediction.  Including the nine charged-fermion masses and the four CKM parameters, the total count of independent flavor parameters is reduced from twenty to nineteen.

Numerically, the model predicts the leptonic CP phase $\dcp = 230.6^\circ \pm 8.4^\circ$, in good agreement with the current global fits. The absolute neutrino masses obtained, $m_1 \approx 1.21 \times 10^{-5}$ eV, $m_2 \approx 8.68 \times 10^{-3}$ eV, $m_3 \approx 5.01 \times 10^{-2}$ eV, are fully consistent with oscillation data, and the flavor-projected masses $\langle m_{\nu_e}\rangle$, $\langle m_{\nu_\mu}\rangle$, $\langle m_{\nu_\tau}\rangle$ lie well within the limits set by the KATRIN experiment. The framework also excludes the inverted mass ordering, since the condition $R(\dcp) = \Rexp$ admits no solution for the IO CP phase in the full interval $[0^\circ,360^\circ]$, as verified with the NuFIT~6.1 best-fit values; this provides an additional testable signature.

By linking the neutrino mass mechanism to the known masses of quarks and charged leptons, and by reducing the number of free parameters while making a definite prediction for $\dcp$, the twisted basis offers a more economical and predictive perspective on the flavor sector of the Standard Model.

\appendix

\section{Linear Algebra Topics}
\label{sec:A.Lin}

Originally published in\cite{marco1992}.

\subsection{Theorem}

\begin{theorem}[Diagonalization of Product of Hermitian Matrices]
If $A$ is Hermitian positive-definite and $B$ is Hermitian, then $AB$ has real eigenvalues and can be diagonalized as
\begin{equation}
\Omega^{-1} (AB) \Omega = \Lambda \equiv \operatorname{diag}(\lambda_i), \quad \Omega\Omega^\dagger = A.
\end{equation}
\label{theorem1}
\end{theorem}

\begin{proof}
The proof requires two lemmas and two corollaries.
	
\begin{lemma}[Hermitian Square Root]
Let $A$ be a Hermitian matrix with real, non-negative eigenvalues. Then $A^{1/2}$ is Hermitian.
\label{lemma2}
\end{lemma}
\begin{proof}
Since $A$ is Hermitian, there exists a unitary matrix $U$ such that:
\begin{equation}
U^\dag{}AU = D, 
\end{equation}
where $D$ is a real diagonal matrix. Then,
\begin{equation}
A^{1/2}=UD^{1/2}U^\dag. 
\end{equation}
Thus, the Hermitian adjoint satisfies
\begin{equation}
{(A^{1/2})}^\dag=U(D^{1/2})^\dag{}U^\dag. 
\end{equation}
Therefore, for $A^{1/2}$ to be Hermitian, it is necessary that
\begin{equation}
{(D^{1/2})}^\dag=D^{1/2}, 
\end{equation}
confirming $A^{1/2}$ is Hermitian, which shows that the eigenvalues of $A$ must be real and non-negative.
\end{proof}
	
\begin{lemma}[Reality of Eigenvalues]
Let $A$ be Hermitian with all its eigenvalues positive-definite
and $B$ Hermitian. Then $AB$ has real eigenvalues.
\label{lemma3}
\end{lemma}
\begin{proof}
Since $\det(A)\neq0$, the matrix $AB$ can be rewritten as
\begin{equation}
AB=A^{1/2}L(A^{1/2})^{-1},  \label{simil}
\end{equation}
where
\begin{equation}           
L\;\equiv\;A^{1/2}BA^{1/2}. \label{L def}
\end{equation}
Since $A^{1/2}$ and $B$ are Hermitian, $L$ is also Hermitian. Lemma~\ref{lemma2} states that $A^{1/2}$ being Hermitian implies $A$ itself is Hermitian with non-negative eigenvalues. Combining this property of $A$ with $\det(A) \neq 0$ satisfies the conditions required for $A$ in lemma~\ref{lemma3}. Furthermore, Eq.~\eqref{simil} shows that $L$ and $AB$ are similar matrices; consequently, they share the same eigenvalues. As $L$ has real eigenvalues, $AB$ must also have real eigenvalues, despite not being Hermitian. 
\end{proof}
	
\begin{corollary}
[Recasting the system of matrix equations]
\label{setting}
\end{corollary}
	
\noindent{}Given that the conditions for $A$ and $B$ in lemma~\ref{lemma3} are satisfied, there exists a unitary matrix $\mathcal{U}$ that diagonalizes the Hermitian $L$. That is, 
\begin{equation}
\def\arraystretch{1.5}
\begin{cases}
\mathcal{U}^\dag\mathcal{U}=\I \\
\mathcal{U}^\dag{}L\,\mathcal{U}=\Lambda,
\end{cases}  \label{unit}
\end{equation}                   
where $\I$ is the identity and $\Lambda$ is diagonal real.
Since
\begin{equation}
AB=A^{1/2}L{(A^{1/2})}^{-1}=
A^{1/2}\mathcal{U}\;\Lambda\;\mathcal{U}^\dag(A^{1/2})^{-1}=
\Omega\;\Lambda\;\Omega^{-1}, \label{AB}
\end{equation}
where
\begin{equation}
\Omega\;\equiv\;A^{1/2}\mathcal{U}. \label{omega}
\end{equation}
From Eq.~\eqref{omega}, we find that
\begin{equation}
\Omega\Omega^\dag=A.  \label{omega con}
\end{equation}
Thus, the matrix $\Omega$ is non-unitary. Note that Eq.~\eqref{omega con} is equivalent to the first equation of the system~\eqref{unit}, and Eq.~\eqref{AB} is equivalent to the second. Therefore, we can recast the system~\eqref{unit} as
\begin{equation}
\def\arraystretch{1.5}
\begin{cases}        
\Omega\Omega^\dag=A \\
\Omega^{-1}AB\,\Omega=\Lambda.
\end{cases}                 \label{n-unit1}
\end{equation}                 
This yields a non-unitary diagonalization equivalent to the unitary diagonalization in system~\eqref{unit}. 
	
\begin{corollary}
[Solution to the eigenvalue problem of $AB$]
\label{eigenvalueproblem}
\end{corollary}
	
This problem consists of finding $\Omega$ and $\Lambda$ from the known matrices $A$ and $B$ with the conditions of the lemma~\ref{lemma3} satisfied.
Assume that $A$ and $B$ are $N\!\times\!N$, the diagonal $\Lambda$ has eigenvectors,
\begin{equation}
u_1 = \begin{pmatrix} 1 \\ 0 \\ \vdots \\ 0 \end{pmatrix}, \quad
u_2 = \begin{pmatrix} 0 \\ 1 \\ \vdots \\ 0 \end{pmatrix}, \quad
\dots, \quad
u_N = \begin{pmatrix} 0 \\ 0 \\ \vdots \\ 1 \end{pmatrix},
\end{equation}
such that
\begin{equation}
\Lambda\,u_i=\lambda_i\,u_i.    \label{trivialeigen}
\end{equation}
As is standard for unitary matrices, the eigenvectors $u_i$ satisfy the orthonormality and completeness relations: 
\begin{equation}
u_i^\dag{}u_j=\delta_{ij}, \label{orto1}
\end{equation}
and
\begin{equation}
\sum_i{}u_i{}u_i^\dag=\I.  \label{compl1}
\end{equation}
Using the second equation of the system~\eqref{n-unit1} to eliminate $\Lambda$ from Eq.~\eqref{trivialeigen}, we obtain
\begin{equation}
\Omega^{-1}AB\,\Omega{}u_i=\lambda_i\,u_i. 
\end{equation}
Left-multiplying both sides by $\Omega$ yields the eigenvalue equation,
\begin{equation}
AB\;\omega_i=\lambda_i\;\omega_i,
\label{autov}
\end{equation}
where the eigenvectors of $AB$ are given by
\begin{equation}
\omega_i=\Omega{}u_i.
\label{vet}
\end{equation}
This explicitly shows that $AB$ and $\Lambda$ share the same eigenvalues $\lambda_i$ ($i=1,\ldots{},N$), which can be found by solving the characteristic equation,
\begin{equation}
\det(AB-\lambda_i{}\I)=0. 
\end{equation}
Using Eqs.~\eqref{vet} and \eqref{compl1}, we find
\begin{equation}
\sum_i\omega_i\omega_i^\dag=
\Omega\Big(\sum_i{}u_i{}u_i^\dag\Big)\Omega^\dag=
\Omega\Omega^\dag. 
\end{equation}
From the first equation in the system~\eqref{n-unit1}, we obtain the completeness relation
\begin{equation}
\sum_i\omega_i\omega_i^\dag=A.  
\label{compl2}
\end{equation}
By substituting the eigenvalues $\lambda_i$ into Eq.~\eqref{autov}, we find the eigenvectors $\omega_i$, which are subsequently normalized using Eq.~\eqref{compl2}. The normalization constants are determined by solving the system of equations
\begin{equation}
\sum_{i=1}^N c_i \, \omega_i \omega_i^\dagger = A
\label{eq:norm}
\end{equation}
for the positive coefficients $c_i$. In practice, this system can be solved using only the $N$ real equations extracted from the diagonal of the matrix equation~\eqref{eq:norm}. The eigenvectors are then rescaled as $\omega_i \to \sqrt{c_i} \, \omega_i$.

Once the eigenvectors $\omega_i$ are normalized, the remaining task is to build $\Omega$. Given that each column is obtained from Eq.~\eqref{vet} as
\begin{equation}\label{eq:omegacol}
\omega_i = \Omega u_i = \begin{pmatrix} \Omega_{1i} \\ \Omega_{2i} \\ \vdots \\ \Omega_{Ni} \end{pmatrix},
\end{equation}
the matrix $\Omega$ is constructed as
\begin{equation}\label{eq:omegaMatrix}
\Omega = \begin{pmatrix} | & | & & | \\ \omega_1 & \omega_2 & \cdots & \omega_N \\ | & | & & | \end{pmatrix}.
\end{equation}
	
The main result follows directly from these corollaries. The diagonalization is achieved through $\Omega$, and the eigenvalues are real by lemma \ref{lemma3}.
\end{proof}

\section{Analytical solution of the neutrino mass cubic equation}
\label{sec:Cardano}

In the twisted-basis extension of the Standard Model, the neutrino masses are the eigenvalues of the matrix
\begin{equation}
H = K^{-1} M,
\label{eq:defH}
\end{equation}
where
\begin{equation}
K = \operatorname{diag}\!\left(\frac{m_u}{m_d},\, \frac{m_c}{m_s},\, \frac{m_t}{m_b}\right)
\label{eq:Kdef}
\end{equation}
is the dimensionless quark kinetic matrix and
\begin{equation}
M = \alpha_\nu\, U\, m_\E\, U^\dagger,
\qquad m_\E = \operatorname{diag}(m_e, m_\mu, m_\tau).
\label{eq:Mdef}
\end{equation}
Here $U$ is the standard PMNS unitary matrix, which depends on the three mixing angles $\theta_{12}, \theta_{13}, \theta_{23}$ and on the CP-violating phase $\dcp$. The constant $\alpha_\nu$ is a dimensionless overall scale parameter. Among these parameters, only $\dcp$ and $\anu$ are free parameters, while the remaining ones are fixed inputs.

The characteristic equation for $H$ is a cubic,
\begin{equation}
\det(H - m \I_3) = 0 \quad\Longrightarrow\quad m^3 - T m^2 + S m - D = 0,
\label{eq:cubic}
\end{equation}
with the invariants
\begin{equation}
T = \operatorname{Tr}(H),\qquad
S = \frac{1}{2}\bigl(T^2 - \operatorname{Tr}(H^2)\bigr),\qquad
D = \det(H).
\end{equation}
For the experimental best-fit values of the quark masses, charged-lepton masses and mixing angles, the three roots are real and distinct. The standard Cardano formula with complex radicals produces excessively long expressions. We therefore employ the trigonometric (Viète) form to obtain compact real analytical expressions.

\subsection*{Reduction to an $\alpha_\nu$-independent cubic equation}
\label{sec:reduction}

Because $M = \alpha_\nu\, U m_\E U^\dagger$, the matrix $H$ factorizes trivially:
\begin{equation}
H = \alpha_\nu\, h,\qquad h = K^{-1}\, U\, m_\E\, U^\dagger.
\label{eq:hdef}
\end{equation}
The matrix $h$ carries dimensions of mass (GeV). Its invariants,
\begin{equation}
t = \operatorname{Tr}(h),\quad
s = \frac{1}{2}\bigl(t^2 - \operatorname{Tr}(h^2)\bigr),\quad
d = \det(h),
\label{eq:invariants}
\end{equation}
are related to $T, S, D$ by
\begin{equation}
T = \alpha_\nu t,\qquad S = \alpha_\nu^2 s,\qquad D = \alpha_\nu^3 d.
\end{equation}
Substituting these relations into the cubic equation given in Eq.~\eqref{eq:cubic} and setting $m = \alpha_\nu \lambda$, we obtain a cubic equation that is completely independent of $\alpha_\nu$:
\begin{equation}
\lambda^3 - t \lambda^2 + s \lambda - d = 0,
\label{eq:cubic_reduced}
\end{equation}
whose coefficients depend exclusively on the mixing angles and on the CP phase $\dcp$ (through $\cos\dcp$). The physical neutrino masses are then recovered as
\begin{equation}
m_i = \alpha_\nu\, \lambda_i(\dcp).
\label{eq:mass_scale}
\end{equation}
When computing the mass splitting ratio $R(\dcp) = \Delta m_{31}^2 / \Delta m_{21}^2$, the scalar factor $\anu^2$ cancels out identically:
\begin{equation}
R(\dcp)=\dfrac{\lambda_3^2(\dcp)-\lambda_1^2(\dcp)}{\lambda_2^2(\dcp)-\lambda_1^2(\dcp)}.	
\end{equation}
As the $\lambda_i(\dcp)$ ($i=1,2,3$) are three distinct real roots of Eq.~\eqref{eq:cubic_reduced}, $R(\dcp)$ is invertible if and only if all roots remain distinct, with the invariants in Eq.~\eqref{eq:invariants} satisfying the relations $s \neq d/t$ and $t \neq 0$, which are indeed satisfied. Consequently, the equation $R(\dcp) = \Rexp$ uniquely determines $\dcp$. The value of $\Rexp$ is obtained directly from the data given in Table~\ref{PMNS_table} in the main text.

\section*{Acknowledgments}

\noindent M.A. De Andrade thanks José Abdalla Helayël-Neto for introducing him to the topic of massive neutrinos and for his valuable assistance in the development of the theorem in this work.

\vspace{5mm}

\noindent E.V. Corrêa Silva thanks Universidade
do Estado do Rio de Janeiro (UERJ) for the Prociência grant.

\vspace{5mm}

\noindent This work is partially supported by FAPERJ and CNPq
(Brazilian Research Agencies).

\vspace{5mm}

\noindent\textbf{Author Contributions.} All authors contributed equally to the elaboration
of this manuscript, from its conception to its final form.

\bibliographystyle{elsarticle-num}

\end{document}